# Broadband Energy Harvesting Using a Metamaterial Resonator Embedded With Non-Foster Impedance Circuitry

Guoqing Fu and Sameer Sonkusale[a)]

*NanoLab, Department of Electrical and Computer Engineering, Tufts University, 161 College Ave,*

*Medford, Massachusetts, 02155, USA*

Radio Frequency Identification (RFID) and implantable biomedical devices need efficient power and data transfer with very low profile antennas. We propose a low profile electrically small antenna for near-field wireless power and data telemetry employing a metamaterial Split Ring Resonator (SRR) antenna. SRRs can be designed for operation over wide frequencies from RF to visible. However, they are inherently narrowband making them sensitive to component mismatch with respect to external transmit antenna. Here we propose an embedding of a non-foster impedance circuitry into the metamaterial SRR structure that imparts conjugate negative complex impedance to this resonator antenna thereby increasing the effective bandwidth and thus overcoming the fundamental limit for efficient signal coupling. We demonstrate the concept through extensive numerical simulations and a prototype system at the board level using discrete off-the-shelf components and printed circuit SRR antenna at 500 MHz. We show that the power transfer between SRR receive antenna and the external transmit loop antenna is broadened by almost 400 MHz which corresponds to increase in $\Delta f/f_C$ from 0.49 to 1.65, before and after non-foster circuit activation.

## I. INTRODUCTION

Wireless and wearable implantable biomedical devices are promising for continuous real-time measurement of underlying physiological signals such as electrocardiography in the heart[1], electromyography in the muscle[2], action potentials in brain[3] etc. In these biomedical devices, wireless power and data telemetry plays a key function that prolongs their operational life by promoting communication based on backscattering that can run without a battery or on single rechargeable battery.

Magneto-inductive telemetry is widely used in these devices for such wireless power and data transfer. The typical inductive telemetry core consists of a pair of antenna/coils placed coaxially in space, one inside the device and one placed externally as part of the reader/interrogator. The external coil typically transmits data which is also harvested for its power and regulated to power up the implant circuitry. The antennas used for such telemetry systems are either loop wire coils or printed spiral coils. One of the most important design criteria is to maximize the coupling coefficient between external and implanted coils, which affects the power/data transfer efficiency significantly.

Recently, there has been an effort utilizing metamaterials for improved wireless energy transfer. Metamaterial are artificial structures that can exhibit exotic electromagnetic properties such as negative index of refraction, which was initially theorized by Veselago[4] and experimentally proved by Smith *et al.* [5]. Split Ring Resonator (SRR) forms the basis of many metamaterial structures and has been studied extensively in the literature. One of the advantage for SRR and other metamaterial resonator structures is that it is easy to adapt its unit cell's geometry sizes[6] over wide frequencies from RF to visible.

SRR can exhibit a strong magnetic resonance to EM wave[7] with high quality factor (Q) at resonance. This makes it an excellent candidate for near-field resonant power/data transfer. The SRR also enables sub-wavelength focusing of the incident energy at resonance in the capacitive gap of the metamaterial which could be harvested for energy/power. Compared to conventional antenna/coil system for power telemetry, SRR provides a compact and low profile geometry due to an integrated antenna and resonator function built into the structure[8].

Coupled Mode Theory (CMT) [9] predicts that when the resonances of both coils are matched, under the condition of the "strongly coupled regime", it results in maximal energy transfer between coils. However, if the resonance of either of coils deviates from one another, the power transfer efficiency drops sharply. Since the biological environment is expected to vary due to physiological changes, it is very hard to match their resonances. This issue will be especially critical for any metamaterial-based energy harvesters[8, 10] due to their sharp narrowband resonance, thus automatic tuning of resonances will be necessary. This can be done for conventional coils with tunable varactors[11-12]. However the tuning needs precision and adds complexity in the system and moreover may be difficult with motion artifacts. In this paper, we propose an embedding of a non-foster impedance circuitry inside the SRR metamaterial to achieve optimal power transfer over a broadband condition without the need for any tuning.

Briefly, the approach relies on overcoming the well-known Foster's reactance theorem which states that the reactance for both passive inductor and capacitor increases with frequency[13]. In contrast, the non-foster elements are essentially negative inductors and capacitors that exhibit negative slope of reactance, which can be realized by active negative impedance converters (NICs) [13]. As is illustrated in

---

[a)] Electronic mail: sameer@ece.tufts.edu

Fig. 1, traditional LC matching network containing passive elements can only achieve zero net reactance at certain frequency, while introducing negative inductors and capacitors, non-foster networks can cancel antenna's inherent passive reactance over a wide frequency range thus achieving broadband response. Early in 1970s, Poggio and Mayes first applied non-foster circuit for bandwidth extension of dipole antennas[14]. Matching networks containing non-foster elements are not constrained by the trade-off between bandwidth and efficiency as in passive matching networks according to the Fano-Youla gain-bandwidth theory[15-16].

We employ this non-foster impedance matching to a single split ring resonator on the implant side for power telemetry. This guarantees "strongly coupled regime" over a broad frequency range overcoming the narrowband limitation of SRR resonator antenna. This paper presents detailed numerical simulation results, and the design of such an SRR antenna for the implant (a conventional wound coil is used for an external antenna). The SRR antenna is designed for self-resonance of 500 MHz and contains a non-foster circuitry embedded within its structure. Experimental results are also presented for a prototype which agrees with numerical results. Results show that non-foster circuitry can improve bandwidth and thus the robustness of the wireless link against resonance mismatch from environmental, biological and motion artifacts. Beyond the specific application for energy harvesting, the approach of using non-foster impedance termination to broaden the bandwidth has validity in other metamaterials and antenna applications.

## II. PRINCIPLE OF APPROACH

When a magnetic excitation is applied perpendicular to the SRR plane, the SRR behaves as an RLC resonator driven by an external electromotive force[17]. The SRR's self-resonance is determined by the equivalent passive reactance elements (inductance L primarily depends on the length of the coil and capacitance C depends on the dimensions of the split gap). Considering the non-foster matching concept illustrated in Fig. 1, these inherent passive elements of the SRR can be compensated to achieve zero net reactance over a wide frequency range.

Fig. 2 shows the equivalent circuit model of the SRR antenna with non-foster circuitry termination. Assume a source voltage $V_S$ is applied to the antenna, $R_S$, $L_S$ and $C_S$ is the inherent resistance (including both loss resistance $R_{loss}$ and resistance $R_{rad}$ which is the free space impedance for EM radiation or source resistance in near field excitation), inductance and capacitance of the SRR antenna. The non-foster block introduces non-foster elements $-L_N$ and $-C_N$ to the circuitry. $V_L$ is the voltage acquired by the load resistance $R_L$.

Analyzing this circuit, one can write the equation as

$$V_S - i\left(R_S + s \cdot L_S + \frac{1}{s \cdot C_S}\right) - i\left[s \cdot (-L_N) + \frac{1}{s \cdot (-C_N)}\right] - iR_L = 0$$

$$V_L = iR_L$$

Rearranging the above two equations,

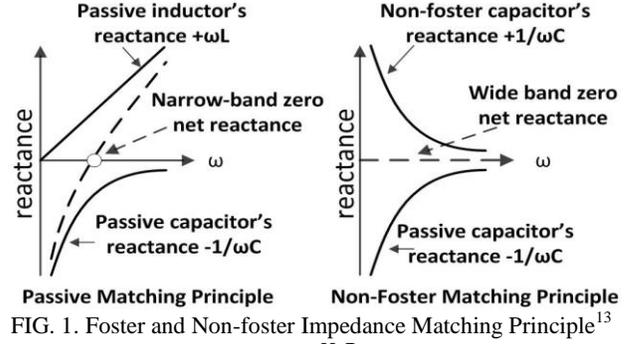

FIG. 1. Foster and Non-foster Impedance Matching Principle[13]

$$V_L = \frac{V_S R_L}{R_S + R_L + (s \cdot L_S - s \cdot L_N) + \left(\frac{1}{s \cdot C_S} - \frac{1}{s \cdot C_N}\right)}$$

By choosing $L_N$ to be equaled with $L_S$ and $C_N$ to be equaled with $C_S$ in a wide frequency range, we get

$$V_{L,max} = \frac{V_S R_L}{R_S + R_L}$$

Thus the reactance of the entire circuitry can be cancelled and the magnitude of voltage obtained across the load impedance $R_L$ is independent of the frequency.

Recent works[18-19] have applied non-foster impedance termination to achieve wide bandwidth for electrically small antennas. The goal in these works was to broaden the bandwidth over which the antenna return loss is minimal for RF transmission/reception. In this paper, we utilize non-foster compensation for wireless power telemetry over broadband.

## III. SRR DESIGN AND SIMULATION

Poon et al.[20] proved that the optimal frequency for power telemetry into biological tissue is in the sub-GHz region. Moreover, a SRR with a small inherent resistance is required for efficient power/data transmission to a load. Based on these observations, a SRR unit cell (TABLE. 1) with 500 MHz self-resonance is designed in CST Microwave Studio (http://www.cst.com) and fabricated on FR4 PCB board. The exact values of $L_S$, $C_S$ and $R_S$ are 33nH, 3pF and 3ohm respectively at frequencies around 500 MHz, which is simulated and extracted using finite difference time domain (FDTD) solver in CST Microwave studio.

A key design consideration is to find a location where to embed the non-foster circuitry that will generate negative reactance, which we found to be a break in the middle of the branch of SRR as shown in Fig. 2. The break in the coil is small enough that it does not affect the primary resonance, but adds a higher order mode at much higher frequency. A similar procedure of breaking the SRR for active circuit embedding was proposed in our earlier work[21], where the goal was to achieve loss compensation in metamaterials.

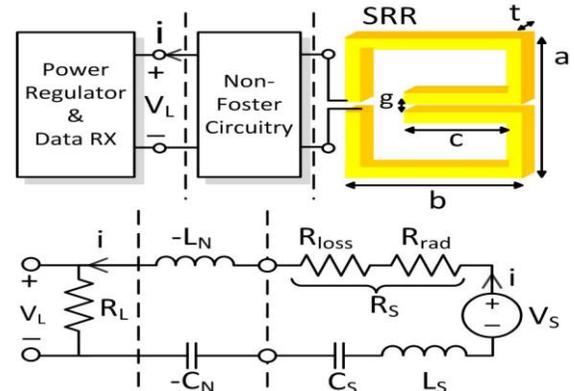

FIG. 2. Equivalent circuit model of SRR antenna with non-foster impedance termination

TABLE I.    SRR UNIT CELL PARAMETERS

| a(mm) | b(mm) | c(mm) | g(mm) | t(mm) |
|---|---|---|---|---|
| 20 | 20 | 14 | 0.2 | 0.03 |

The loop antenna is placed 5mm above the SRR structure and is aligned coaxially so that the magnetic field for excitation is perpendicular to both the SRR and the loop antenna plane[22]. The entire wireless telemetry system including a loop antenna and a SRR unit with the break termination designed in CST is then extracted as a two-port network and fetched to Advanced Design System (ADS) to apply circuitry elements connections as shown in Fig. 3. The primary port connected to loop wire provides power while the secondary port is connected at the SRR terminal to measure received power level and the SRR resonance. Both ports have 50ohm impedance to represent the VNA ports used in experiment as indicated in section V.

Applying CST-ADS co-simulation and sweeping frequency from 0 to 1GHz, the dotted plots in Fig. 4 (S22 and S21) represent the SRR's self-resonance and received power level respectively. In comparison, the solid plots (S44 and S43) are those of the SRR connected with ideal negative inductor and capacitor at Port 4. From the plots we can see that with ideal non-foster elements, the SRR has a 100MHz broadened -10 dB bandwidth that spans from 80MHz to 180MHz. In terms of received power level, with reference to the same -15dB level, there is also significant improvement that spans from 280MHz to 900MHz, compared with that of the uncompensated SRR, which spans from 400MHz to 700MHz. These results indicate the merit of non-foster compensation for power/data transmission using SRR antenna.

## IV. NON-FOSTER CIRCUITRY

There are several versions of non-foster element generation circuitry, among which the Linvill OCS (Open Circuit Stable) circuit has been studied for negative capacitance generation[13]. However, we need to compensate both L and C of SRR simultaneously. The basic Linvill OCS circuit and the designed circuitry connected with SRR is shown in Fig. 5. In our realization, the bipolar transistors of the original Linvill OCS are replaced by pHEMT transistors for their high mobility and low parasitics at high frequency. The Linvill circuit is designed with 10mA bias current and 2V supply voltage. With conventional inductor L and capacitor C connected at drain terminals of the cross coupled transistor pair, the effective non-foster impedance is generated at the source (output) side; this is connected in series with the SRR unit cell at the break location. To indicate the performance of this non-foster circuitry, the effective negative inductances and capacitances with particular load are plotted in supplementary section (Fig. S1 and Fig. S2). Instead of ideal non-foster elements, both the generated negative inductance and capacitance actually vary with frequency; however there is a wide frequency band over which negative values of impedances are stably generated. One can also see that the negative impedance can be tuned by varying the values of loaded conventional inductors and capacitors.

## V. EXPERIMENT

The experiment setup is shown in Fig. 6. The fabricated non-foster PCB board is 4.5cm by 6.6cm. The power for transmission is fed by a SMA connector at the end of loop wire to Port 1 of Vector Network Analyzer (VNA). The SRR with non-foster termination is fed to Port 2 of VNA to measure the self-resonance and received power level before and after compensation. The experimental results are indicated in Fig. 7.

As indicated by the dash curve in Fig. 7a, the S11 measurement at -10dB is narrowband (around 10MHz) at center frequency of 490MHz before non-foster activation. Terminated with on-board non-foster PCB board, there is broadening of resonance bandwidth (to 40MHz) from 450MHz to 490MHz (solid curve). The addition of non-foster circuitry introduces parasitic elements and creates another mode from 220MHz to 300MHz, however it has low S21 which is a measure of received power (Fig. 7b). Received power level without non-foster termination (dash curve) has a bandwidth of 90MHz from 450MHz to 540MHz at -20dB which is significantly improved to 470MHz (spanning 240MHz to 710MHz) (Fig. 7b). This corresponds to an increase of $\Delta f/f_C$, a measure of broadening from 0.49 to 1.65. This shows that the SRR antenna with non-foster circuitry termination will able to receive signal without the need to match resonance to the antenna of power/data transmitter.

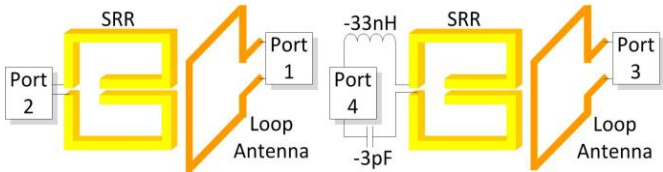

Fig. 3. Wireless system consisting loop antenna and SRR (left) and the one applying ideal negative impedance compensation (right)

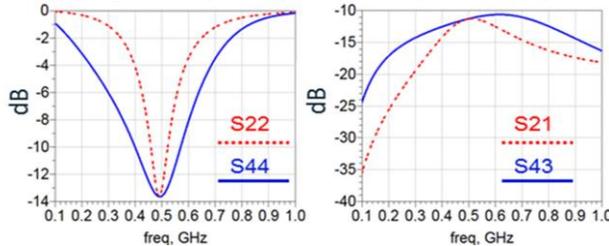

Fig. 4. SRR resonance (left) and received power (right) broadening simulation results

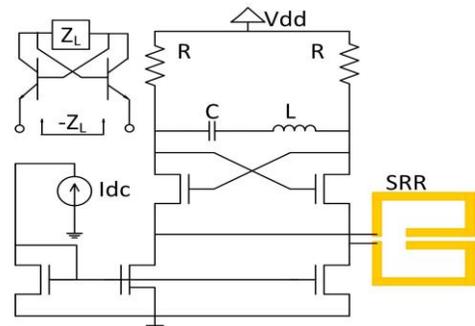

FIG. 5. Basic Linvill OCS circuit and negative LC generation circuitry using pHEMT

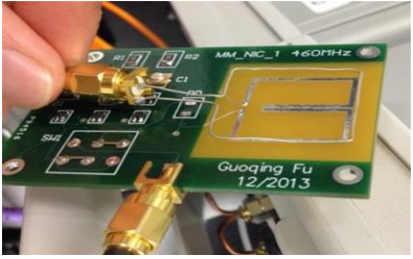

FIG. 6. Measurement Setup

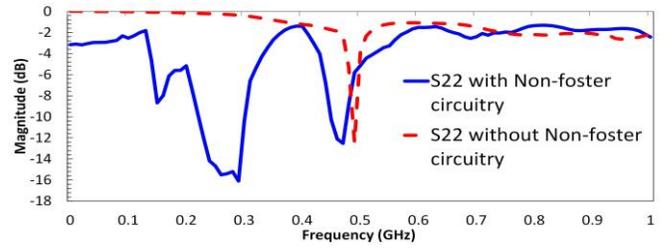

FIG. 7a. S22 (SRR resonance) Comparison

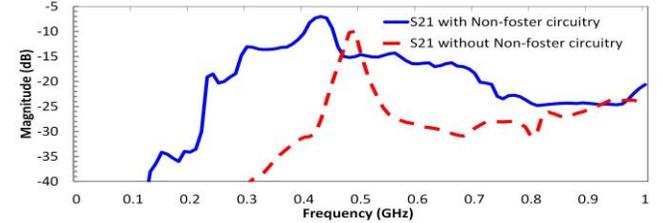

FIG. 7b. S21 (Received power) Comparison

## VI. DISSCUSSION

As indicated by experimental results, embedding non-foster impedance circuitry has great merits in broadening the response of metamaterial structures. Another important observation is that the Linvill OCS circuitry uses positive feedback to generate negative impedance and could be unstable. However recent works have proven that such bandwidth broadening technique is capable of yielding a predictably stable, low noise and high quality performance[23-25].

While the approach was been applied to metamaterial based energy harvester, this network can also be applied to conventional coil structure used in power telemetry. Other metamaterial applications where wide-band absorption or transmission is necessary can also benefit from the broadening effect. Moreover, the SRR metamaterial antenna with non-foster termination proposed in this paper can be widely used as a wide-band front-end module in wireless implantable devices for different applications.

## VII. SUMMARY

This paper presented a design of an electrically small Split Ring Resonator (SRR) metamaterial antenna for resonant power/data telemetry in biomedical and RFID applications. SRR based antenna provides inherent high quality factor and narrowband response that while ideal for resonant power transfer, is quite sensitive to the biological environment and motion artifacts causing loss of efficiency of wireless energy transfer. To combat this issue, we showed the embedding of a non-foster circuitry to cancel the reactive inductance and capacitance of the SRR to effectively broaden its bandwidth. As a result, the SRR's received power level are broadened. Both numerical results and experimental data verify the approach. This topology can be used to improve wireless power/data telemetry, in biomedical and RFID applications. And the use of non-foster impedance circuitry can be employed for other metamaterial and electrically small antenna applications.